# Particle-hole instabilities in photonic time-varying systems

JOÃO C. SERRA[1,#], EMANUELE GALIFFI[2], PALOMA A. HUIDOBRO[1,3], J. B. PENDRY[4], AND MÁRIO G. SILVEIRINHA[1,*]

[1]*University of Lisbon – Instituto Superior Técnico and Instituto de Telecomunicações, Avenida Rovisco Pais, 1, 1049-001 Lisboa, Portugal*
[2]*Advanced Science Research Center, City University of New York, 85 St. Nicholas Terrace, 10031 New York, NY, USA*
[3]*Departamento de Física Teórica de la Materia Condensada and Condensed Matter Physics Center (IFIMAC), Universidad Autónoma de Madrid, E-28049 Madrid, Spain USA*
[4]*The Blackett Laboratory, Department of Physics, Imperial College London, London, SW7 2AZ, UK*
[#] *joao.serra@lx.it.pt*
[*] *mario.silveirinha@tecnico.ulisboa.pt*

**Abstract:** Photonic systems with time-varying modulations have attracted considerable attention as they allow for the design of non-reciprocal devices without the need for an external magnetic bias. Unlike time-invariant systems, such modulations couple modes with different frequencies. Here, we discuss how this coupling and particle-hole symmetry may lead to the resonant interaction of positive and negative frequency oscillators. To illustrate this idea, we analyze a dispersive spacetime crystal described by a Drude-Lorentz model with a traveling-wave modulation. Our findings demonstrate that the interaction between positive and negative frequency bands can induce parametric instabilities under certain conditions, stemming from the interplay between dispersion and spacetime modulations. In particular, we find that material dispersion creates the conditions for the formation of instabilities for arbitrarily small modulations speeds in the absence of dissipative channels.

## 1. Introduction

Conventional photonic crystals are material structures whose electromagnetic response is periodic in space [1, 2]. They were originally introduced as a proposal to inhibit the spontaneous emission in optical cavities [3] and as a new mechanism for strong Anderson localization of photons [4]. By controlling the shape, the size, and the materials in the unit cell, one may tailor the resonant light-matter interactions and mold the flow of light [5-11].

Nowadays, motivated by recent technological advancements, the concept of photonic crystal has been broadened to encompass platforms with periodic time-varying modulations [12-14]. These systems enable the design of non-reciprocal devices without an external magnetic bias [15-20], as well as to explore new and exotic wave phenomena [21-31].

Due to their inherent active nature, time-varying systems can exhibit loss or gain [32-34]. In fact, systems with parameters that undergo a periodic modulation over time are known to present instabilities with exponential growth, a phenomenon known as parametric resonance [35-38]. Therefore, it is important to clearly understand the physical mechanisms underlying these instabilities, enabling their exploitation in non-Hermitian regimes or their deliberate avoidance, depending on the application.

Here, we discuss how time-varying modulations couple modes with different frequencies and some of their consequences in terms of stability. We prove that the interactions between photonic bands with positive and negative frequencies are analogous to those between particle-states, with a positive mass, and hole-states, with a negative mass, which are known to lead to instabilities within the system [39-41]. As an example, we analyze the stability of a dispersive spacetime crystal described by the Drude-Lorentz model with a traveling-wave modulation [42]. Furthermore, we explain how the combination of dispersion and spacetime modulation naturally gives rise to these particle-hole instabilities.

Parametric instabilities have been studied in the literature mostly in non-dispersive time-variant systems, but also in some spacetime crystals, e.g., [43-45]. Furthermore, instabilities in dispersive systems were analyzed in Ref. [46], where it was shown that the interaction between modes propagating in the same direction may become unstable, provided that the modulation speed exceeds the wave velocity in the (non-dispersive) dielectric background. In contrast, we shall show that particle-hole instabilities may occur for arbitrarily small modulation velocities, in the absence of dissipative channels.

## 2. Coupling between modes with positive and negative frequencies

Periodic modulations in space and/or time within a material structure create interaction channels between the modes of the unperturbed system. The strength and nature of these interactions are controlled by the modulation period. For example, the band gaps of a Bragg mirror result from the destructive interference between two modes whose wavenumbers differ by an integer multiple of $K = 2\pi/a$, with $a$ being the spatial period [47]. Analogously, temporal modulations of period $T$ couple modes whose frequencies differ by an integer multiple of $\Omega = 2\pi/T$. This relationship becomes evident when we express the homogeneous Maxwell's equations of a dispersionless system in the frequency domain:

$$\begin{pmatrix} \mathbf{0} & +i\nabla\times \\ -i\nabla\times & \mathbf{0} \end{pmatrix} \begin{pmatrix} \mathbf{E}(\mathbf{r},\omega) \\ \mathbf{H}(\mathbf{r},\omega) \end{pmatrix} = \omega \sum_{n=-\infty}^{+\infty} \bar{M}_n(\mathbf{r}) \cdot \begin{pmatrix} \mathbf{E}(\mathbf{r},\omega-n\Omega) \\ \mathbf{H}(\mathbf{r},\omega-n\Omega) \end{pmatrix}. \qquad (1)$$

Here, $\bar{M}_n(\mathbf{r})$ is the $n$-th harmonic of the material constitutive matrix $\bar{M}(\mathbf{r},t)$, and $\mathbf{E}(\mathbf{r},\omega)$ and $\mathbf{H}(\mathbf{r},\omega)$ represent the Fourier transforms in time of the electromagnetic fields.

Different from electronic crystals, their photonic counterparts present both positive and negative frequency bands due to the particle-hole symmetry of the electromagnetic spectrum. In fact, as the electromagnetic field is real-valued, its spectrum has mirror symmetry with respect to the line $\text{Re}\{\omega\} = 0$. Interestingly, this spectral symmetry has important implications regarding the stability of time-varying materials.

To illustrate this, let us consider a closed cavity filled with a dielectric material of permittivity $\varepsilon$ that supports a discrete set of modes $\psi_{\pm n}$ with frequencies $\pm\omega_n$ as represented in Fig. 1. Suppose that before the time modulation is switched on ($t<0$) the $n=1$ mode is populated via external driving. By introducing a weak temporal modulation $\varepsilon \to \varepsilon[1+\delta\sin(\Omega t)]$ with $\delta \ll 1$, it is possible to resonantly couple modes whose natural frequencies differ by the modulation frequency $\Omega$ [48, 49].

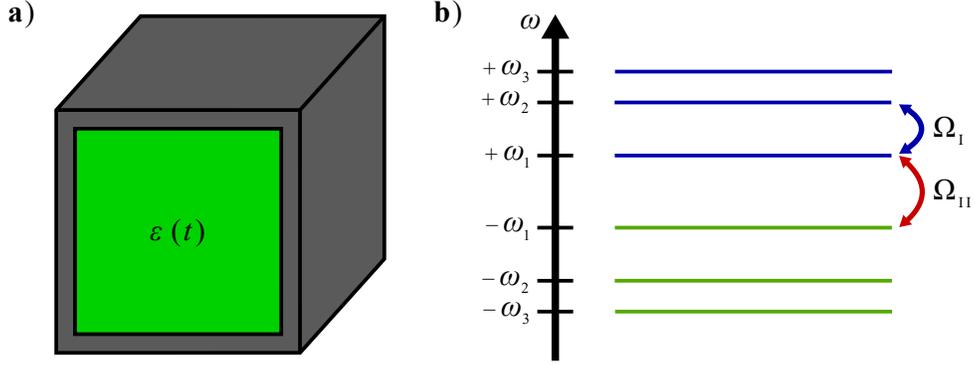

**Fig. 1 (a)** Closed cavity filled with a dielectric material of permittivity $\varepsilon$. **(b)** Discrete spectrum of the cavity modes with positive (blue) and negative (green) frequency bands. We represent the coupling due to a temporal modulation of the permittivity between two different modes in blue and the resonant feedback interaction in red.

First, let us choose $\Omega = \Omega_I = \omega_2 - \omega_1$ so that $\psi_{+1}$ and $\psi_{+2}$ are coupled by the temporal modulation. Initially, all the field energy is stored in the $n=1$ mode, but for $t>0$ the temporal coupling leads to a transfer of energy to the $n=2$ mode. We can describe the overall time evolution of the energies $\mathcal{E}_1$, $\mathcal{E}_2$ stored in the $\psi_{+1}$, $\psi_{+2}$ modes using the simple model

$$\frac{d\mathcal{E}_1}{dt} = -\Gamma_1 \mathcal{E}_1(t) + G_{12}\mathcal{E}_2(t),$$
$$\frac{d\mathcal{E}_2}{dt} = +G_{21}\mathcal{E}_1(t) - \Gamma_2 \mathcal{E}_2(t), \tag{2}$$

where $\Gamma_1$, $\Gamma_2$ are decay rates that account for the intrinsic losses of the system and $G_{12}$, $G_{21}$ determine the power transfer between the coupled modes, which is controlled by the modulation strength $\delta$. The system remains stable provided that $G_{12}G_{21} \leq \Gamma_1\Gamma_2$. A coupling with $G_{12}G_{21} < 0$ describes a resonant interaction where the energy flows back and forth from one mode to the other. On the contrary, a coupling of the type $G_{12}G_{21} > 0$ means that either the modes provide simultaneously power to each other ($G_{12}, G_{21} > 0$), leading to an unstable system, or they absorb energy from one another ($G_{12}, G_{21} < 0$), creating a lossy system. Obviously, none of the latter scenarios is possible in passive conservative systems because energy cannot be created nor destroyed. However, time-varying platforms are active systems, and they always require an external source of energy to change their configuration in time.

Interestingly, when $\Omega = \Omega_{II} = 2\omega_1$, $\psi_{+1}$ and $\psi_{-1}$ become coupled. However, $\psi_{\pm 1}$ are two components of the same mode as the reality condition imposes $\psi_{-1} = \psi_{+1}^*$ and thus the resonant interaction creates a feedback loop:

$$\frac{d\mathcal{E}_1}{dt} = (G_{11} - \Gamma_1)\mathcal{E}_1(t). \tag{3}$$

Now, the system only reaches a steady state if $G_{11} \leq \Gamma_1$. Otherwise, the energy inside the cavity grows exponentially in time, i.e., $\mathcal{E}_1(t) = \mathcal{E}_{t=0}\exp\{(G_{11}-\Gamma_1)t\}$, turning the system unstable until nonlinear effects come into play and the energy saturates. This self-excitation mechanism due to the time modulation is generally known as parametric resonance [35-38]. Importantly, while the coupling between different modes is a second-order process ($G_{12}G_{21} \sim \delta^2$), the feedback loop associated with the self-excitation is a first-order mechanism ($G_{11} \sim \delta$).

The resonant interactions between different modes can couple oscillators with the same frequency sign or with opposite frequency sign (being the self-excitation an example of special interest). In the following, we show that in photonic time-varying systems the interaction between bands with identical frequency sign is usually stable, while the instability condition $G_{12}G_{21} > 0$ corresponds to the coupling of positive and negative frequencies.

## 3. Instabilities in a dispersive spacetime crystal

### 3.1 Dispersive spacetime crystal model

Let us consider the dynamical response of a dispersive spacetime crystal with a traveling-wave modulation, as represented in Fig. 2a. We suppose that the dynamics of the polarization vector $P$ is controlled by a dispersive Drude-Lorentz model with time-varying coefficients [42, 46, 50, 51]:

$$\left[\frac{\partial^2}{\partial t^2} + \Gamma \frac{\partial}{\partial t} + \omega_0^2(x-vt)\right] P(x,t) = \varepsilon_0 \omega_p^2 E(x,t). \tag{4}$$

The plasma frequency $\omega_p$ remains constant over time, while the resonance frequency $\omega_0$ undergoes modulation in both space and time. Here, $\Gamma$ plays the role of a "collision frequency" and models the losses in the system. The electric field ($E$) and the displacement vector ($D$) are related as $D = \varepsilon_0 E + P$. The homogeneous Maxwell's equations for a transverse wave propagating along $x$ can be written as [52, 53]

$$\hat{H}(i\partial_x, x-vt) \cdot \psi(x,t) = i\partial_t \psi \tag{5}$$

with the equivalent Hamiltonian operator (for $v = 0$)

$$\hat{H}(i\partial_x, x) \equiv \begin{pmatrix} 0 & -ic\partial_x & 0 & 0 \\ -ic\partial_x & 0 & +ic\partial_x & 0 \\ 0 & 0 & 0 & i\omega_p \\ i\omega_p & 0 & -i\dfrac{\omega_0^2(x)+\omega_p^2}{\omega_p} & -i\Gamma \end{pmatrix}. \tag{6}$$

The state vector is defined by $\psi = \dfrac{1}{\sqrt{W_0}}\left(\dfrac{D}{\sqrt{\varepsilon_0}} \quad \dfrac{B}{\sqrt{\mu_0}} \quad \dfrac{P}{\sqrt{\varepsilon_0}} \quad \dfrac{J}{\omega_p\sqrt{\varepsilon_0}}\right)^T$, with $B$ the magnetic field and $J = \partial_t P$ the polarization current. The normalization constant is chosen as $W_0 = 1\, J/m^3$. It is known that the material dispersion may play an important role in time-varying platforms [54-56].

For convenience, we write the function $\omega_0^2(x)$ in terms of its mean value and a fluctuation: $\omega_0^2(x) = \bar{\omega}_0^2 + \delta\tilde{\omega}_0^2(x)$. Here, $\tilde{\omega}_0^2$ is a periodic zero-mean valued function that determines the shape of the modulation and $\delta$ controls its strength. For simplicity, we focus on a bi-layer crystal with a unit cell such that $\omega_0$ can assume only two values (see Fig. 2a).

## 3.2 Spacetime band crossings

In the static case ($\delta = 0$), the system is homogeneous in space and time. For now, let us ignore the effect of intrinsic dissipation ($\Gamma = 0$). The impact of losses on the system response will be discussed later. Thus, the eigenmodes of the static system are time-harmonic plane waves, i.e., $\psi(x,t) = \tilde{\psi}\exp\{i(kx - \omega t)\}$ with $\omega$ and $k$ real-valued. Figure 2b shows the corresponding band structure composed of two positive frequency bands (in blue) as well as two negative frequency bands (in green).

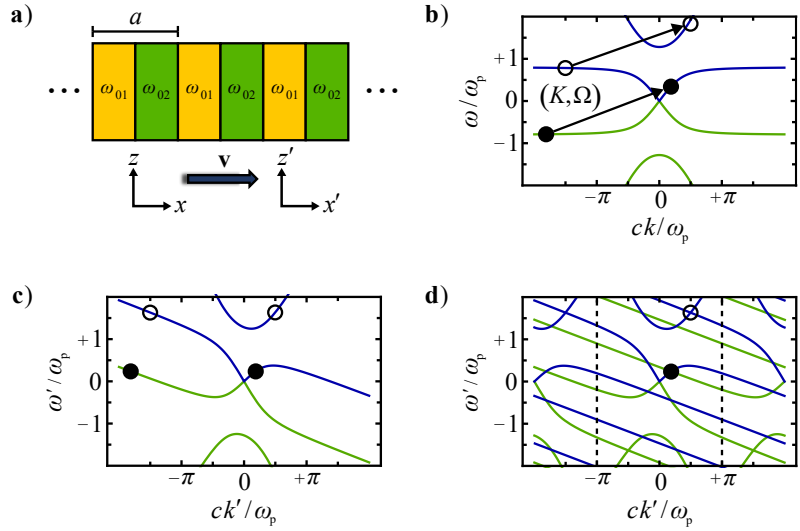

**Fig. 2 (a)** Dispersive spacetime crystal described by a time-varying Drude-Lorentz model with a traveling-wave modulation. **(b)** Band structure of a uniform Drude-Lorentz material in the laboratory frame $(x,t)$ characterized by $\omega_0 = 0.8\omega_p$. We represent the positive/negative frequency bands in blue/green. The black arrows represent the vector $(K,\Omega)$ that controls the modal interactions arising from the spacetime modulation. **(c)** Band structure of the same material in (b) in a Galilean frame described by the coordinate transformation $(x',t') = (x - vt, t)$ with $v = 0.18c$. **(d)** Folding of the band structure in (c) to create a Brillouin zone in the (primed) wavenumber space with $a = c/\omega_p$. The black/white circles identify crossings between two frequency bands of opposite/same sign.

Due to the material dispersion, there are quasi-flat bands with a vanishingly small phase velocity near the resonance frequency $\omega_0$. The spacetime modulation can couple points $(k,\omega)$ of the dispersion diagram linked by the vector $(K,\Omega)$, where $K = 2\pi/a$ and $\Omega = vK$. For example, the white circles in Fig. 2b couple two points in different bands separated by $(1,v)2\pi/a$. Similarly, the black circles couple two modes in the same band through the same vector. Importantly, in the latter case the two black circles are associated with modes with oppositely signed frequencies. In dispersive systems the modal coupling through the vector $(1,v)2\pi/a$ is feasible for arbitrarily small velocities, when the vector $(K,\Omega)$ is almost horizontal, due to the flat bands. We shall see below that this property has interesting implications in terms of the system stability. In contrast, for a non-dispersive dielectric the dispersion diagram is formed by two straight lines and the coupling between modes with oppositely signed frequencies is only possible for a sufficiently large modulation speed.

When the modulation strength is non-trivial ($\delta \neq 0$), the equivalent Hamiltonian in Eq. (5) depends on both space and time. It is possible to suppress the time dependence with a Galilean-type coordinate transformation such that $(x',t') = (x-vt,t)$ [42, 57, 58]. In the new coordinates, the system dynamics reduces to:

$$\hat{H}_{co}(i\partial_{x'}, x') \cdot \psi(x',t') = i\partial_{t'}\psi \tag{7}$$

with the transformed Hamiltonian defined as $\hat{H}_{co}(i\partial_{x'}, x') \equiv \hat{H}(i\partial_{x'}, x') + iv\partial_{x'}$. Thus, in the new coordinates, the effect of the time-modulation can be described in terms of the interaction term $iv\partial_{x'}$. In the limit $\delta \to 0$, the state vector in the primed coordinate system is of the form $\psi(x',t') = \tilde{\psi} \exp\{i(k'x' - \omega't')\}$, with the primed wave number and frequency related to the corresponding parameters in the laboratory frame by a Galilean Doppler shift: $k' = k$ and $\omega' = \omega - vk$. Thus, for a uniform system ($\delta \to 0$), the band structure becomes tilted in the primed coordinates (see Fig. 2c). The envelope $\tilde{\psi}$ is unaffected by the transformation.

The operator $\hat{H}(i\partial_{x'}, x')$ describes the wave dynamics when the modulation is purely spatial. Thus, it models a conventional photonic crystal with spatial period $a$. The effect of the spatial modulation is to fold the dispersion of the uniform system to form a Brillouin zone. This mechanism combined with the effect of the time interaction term $iv\partial_{x'}$ (i.e., with the Doppler shift), leads to the band structure sketched in Fig. 2d, which assumes $\delta = 0^+$. As seen, the spacetime folding creates band crossings between plane wave modes $\psi_{A/B}$ of the uniform system, with wavenumbers $k_{A/B}$ and frequencies $\omega_{A/B}$ in the lab frame that satisfy the following interaction "selection rules":

$$\begin{cases} k_A - N_A 2\pi/a = k_B - N_B 2\pi/a \\ \omega_A - vk_A = \omega_B - vk_B \end{cases}, \quad N_{A/B} \in \mathbb{Z}. \tag{8}$$

It is implicit that $k'_c \equiv k_A - N_A 2\pi/a = k_B - N_B 2\pi/a$ lies in the first Brillouin zone.

Next, we characterize the effect of a weak modulation ($\delta \ll 1$) near a crossing point $(k'_c, \omega'_c)$. To this end, we use perturbation theory to solve the secular problem $\hat{H}_{co}(i\partial_{x'}, x') \cdot \psi = \omega'\psi$. We write $\hat{H}_{co} = \hat{H}_{co,0} + \hat{H}_{pert}$, where $\hat{H}_{co,0}$ is the Hamiltonian for a continuum (evaluated for $\delta = 0$ and a finite modulation speed $v$) and $\hat{H}_{pert}$ is the perturbation due to a finite $\delta$:

$$\hat{H}_{pert}(x') \equiv -i\frac{\delta\tilde{\omega}_0^2(x')}{\omega_p}\begin{pmatrix} 0 & 0 & 0 & 0 \\ 0 & 0 & 0 & 0 \\ 0 & 0 & 0 & 0 \\ 0 & 0 & 1 & 0 \end{pmatrix}. \tag{9}$$

We write the eigenmodes of $\hat{H}_{co}$ in terms of the eigenmodes of $\hat{H}_{co,0}$ (electromagnetic continuum) near the crossing point:

$$\psi(x') \approx \alpha \psi_A(x') + \beta \psi_B(x'). \tag{10}$$

In this basis, $\hat{H}_{co}$ is represented by the matrix:

$$\hat{H}_{co} \to \begin{pmatrix} \omega'_c + \langle \psi_A | \hat{H}_{pert} | \psi_A \rangle & \langle \psi_A | \hat{H}_{pert} | \psi_B \rangle \\ \langle \psi_B | \hat{H}_{pert} | \psi_A \rangle & \omega'_c + \langle \psi_B | \hat{H}_{pert} | \psi_B \rangle \end{pmatrix}, \tag{11}$$

with $\omega'_c = \omega_A - vk_A = \omega_B - vk_B$. We use the weighted inner product $\langle \mathbf{f} | \mathbf{g} \rangle = (1/a) \int_0^a dx' \, \mathbf{f}^*(x') \cdot \bar{M} \cdot \mathbf{g}(x')$ with

$$\bar{M} = \begin{pmatrix} 1 & 0 & -1 & 0 \\ 0 & 1 & 0 & 0 \\ -1 & 0 & 1 + \dfrac{\bar{\omega}_0^2}{\omega_p^2} & 0 \\ 0 & 0 & 0 & 1 \end{pmatrix}, \tag{12}$$

and pick normalized unperturbed modes: $\langle \psi_A | \psi_A \rangle = \langle \psi_B | \psi_B \rangle = 1$. Importantly, the unperturbed Hamiltonian $\hat{H}_{co,0}$ is Hermitian with respect to this weighted inner product and the metric matrix $\bar{M}$ is consistent with the electromagnetic energy density stored in a material with a conventional Drude-Lorentz dispersion [42], i.e.,

$$\langle \psi_n | \psi_n \rangle \sim \frac{1}{2} \varepsilon_0 \partial_\omega [\omega \varepsilon(\omega)] |E_n|^2 + \frac{1}{2} \mu_0 |H_n|^2 \tag{13}$$

with $\varepsilon(\omega) = 1 + \dfrac{\omega_p^2}{\bar{\omega}_0^2 - \omega^2}$. Thus, the secular equation $\hat{H}_{co}(i\partial_{x'}, x') \cdot \psi = \omega' \psi$ reduces to:

$$\begin{pmatrix} \langle \psi_A | \hat{H}_{pert} | \psi_A \rangle & \langle \psi_A | \hat{H}_{pert} | \psi_B \rangle \\ \langle \psi_B | \hat{H}_{pert} | \psi_A \rangle & \langle \psi_B | \hat{H}_{pert} | \psi_B \rangle \end{pmatrix} \begin{pmatrix} \alpha \\ \beta \end{pmatrix} = (\omega' - \omega'_c) \begin{pmatrix} \alpha \\ \beta \end{pmatrix}. \tag{14}$$

As $\psi_A$, $\psi_B$ describe plane waves in a uniform medium it is simple to check that: $\langle \psi_n | \hat{H}_{pert} | \psi_m \rangle = \left\{ \dfrac{\delta \tilde{\omega}_0^2}{\varepsilon_0 \omega_p^2}(x') e^{i(N_m - N_n)\frac{2\pi}{a}x'} \right\}_{av} \omega_n \dfrac{P_n^* P_m}{W_0}$, where $P_n$ is determined by the third component of the state vector $\psi_n$ and $\omega_n$ is the frequency of the unperturbed mode in the unprimed (lab) coordinates. The operator $\{\ \}_{av}$ represents spatial averaging in a unit cell. As $\tilde{\omega}_0^2$ is a function with zero mean-average, it follows that $\langle \psi_A | \hat{H}_{pert} | \psi_A \rangle = \langle \psi_B | \hat{H}_{pert} | \psi_B \rangle = 0$. Furthermore, one has $\langle \psi_A | \hat{H}_{pert} | \psi_B \rangle = \lambda \omega_A$ and $\langle \psi_B | \hat{H}_{pert} | \psi_A \rangle = \lambda^* \omega_B$ with

$$\lambda = \left\{ \frac{\delta\tilde{\omega}_0^2}{\varepsilon_0 \omega_p^2}(x') e^{i(N_B - N_A)\frac{2\pi}{a}x'} \right\}_{av} \frac{P_A^* P_B}{W_0}.$$ Then, from Eq. (14), we find that the perturbed frequencies in the co-moving frame are

$$\omega'_\pm = \omega'_c \pm |\lambda| \sqrt{\omega_A \omega_B}. \tag{15}$$

### 3.3 Particle-hole instabilities

When $\lambda \neq 0$, we can distinguish two different scenarios in Eq. (15). If the band crossing occurs between two unperturbed modes with positive frequencies as represented by the white circle in Fig. 2d, then $\omega_A \omega_B > 0$, and a conventional gap is formed between two real-valued bands. Naturally, the same holds true for the crossing of two negative frequency bands. However, in the case of the black circles, the spacetime modulation couples a positive frequency mode with a negative frequency mode ($\omega_A \omega_B < 0$) and, as a result, the two perturbed modes have complex-valued frequencies: one of them decays with time ($\text{Im}\{\omega'_-\} < 0$), while the other is unstable and grows exponentially in time ($\text{Im}\{\omega'_+\} > 0$). As already mentioned, this second scenario may occur for arbitrarily small modulation velocities in the presence of flat bands due to material dispersion.

Figure 3 shows the relation between the instability strength ($\text{Im}\{\omega'_+\} = |\lambda|\sqrt{-\omega_A \omega_B}$) and the wavenumber $k'_c$ in the co-moving frame for the crossing of the first positive and negative bands. We consider a binary crystal with spatial period $a = c/\omega_p$ characterized by layers with the same thickness and resonance frequencies $\omega_{0,i} = \sqrt{0.64 \pm 0.05}\,\omega_p$ ($i$=1,2). The location of the band crossing is chosen by controlling the modulation velocity ($0.1256 < v/c < 0.2420$). Interestingly, the instability becomes more prominent as the band crossing approaches the Brillouin zone limit $k'_c = \pi/a$. This situation represents precisely a resonant feedback loop as the corresponding unperturbed modes $\psi_A$ and $\psi_B$ are linked by the particle-hole symmetry, i.e., $(k_A, \omega_A) = (-k_B, -\omega_B)$ and $\tilde{\psi}_B = \tilde{\psi}_A^*$. As explained in Sect. 2, this parametric resonance is a first-order phenomenon, implying that it typically exhibits greater strength compared to interactions of the second-order. Note that the particle-hole condition combined with Eq. (8) gives $2\omega_A = n\frac{2\pi}{a}v$, with $n = 1, 2, \ldots$ For large $n$, the particle-hole resonances ($(k_A, \omega_A) = (-k_B, -\omega_B)$) are associated with the velocities $v \approx \frac{\omega_0 a}{n\pi}$, where we used the approximation $\omega_A \approx \omega_0$. In this article, we restrict our attention to the resonances with $n=1$, which provide the strongest modal coupling.

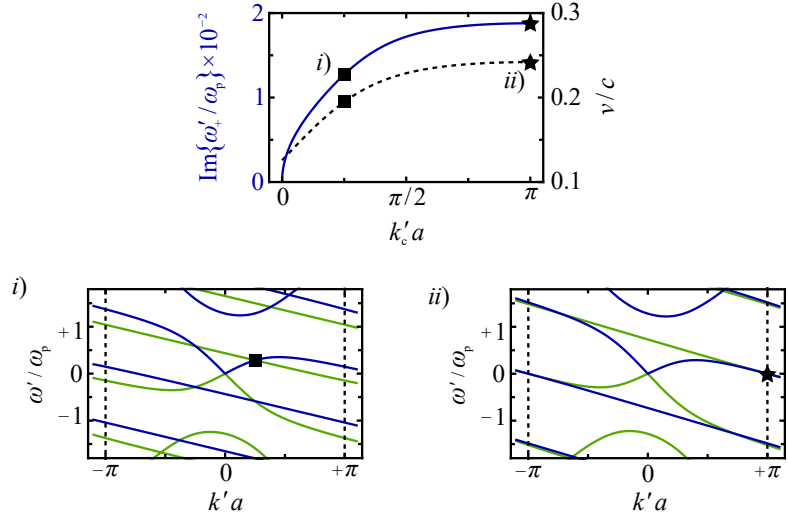

**Fig. 3** Top panel: Instability strength ( $\text{Im}\{\omega'_+\}$, solid blue line) and modulation velocity ( $v$, dashed black line) as a function of the wavenumber $k'_c$ for the crossing between the first positive and negative bands. The parameters of the binary crystal are $a = c/\omega_p$ and $\omega_{0,i} = \sqrt{0.64 \pm 0.05}\,\omega_p$ ($i$=1,2). The instability strength increases as the band crossing approaches the Brillouin zone limit. The labels i) and ii) refer to the points marked in the lower panels. Lower panels: Photonic band structures in the Galilean frame calculated with $\delta = 0^+$ for (i) $v = 0.195c$ and (ii) $v = 0.242c$.

To further explore the model of section 2, let us suppose that some energy is initially stored in mode $\psi_A$. Due to the spacetime modulation, the energy will be weakly coupled to the mode $\psi_B$ according to the following equation:

$$i\frac{d}{dt'}\begin{pmatrix}\alpha\\ \beta\end{pmatrix} = \begin{pmatrix}\omega'_c & \lambda\omega_A\\ \lambda^*\omega_B & \omega'_c\end{pmatrix}\begin{pmatrix}\alpha\\ \beta\end{pmatrix} \tag{16}$$

It is simple to check that the energy dynamics is controlled by:

$$\begin{aligned}\frac{d}{dt'}|\alpha|^2 &= 2\,\text{Re}\{(-i\lambda\omega_A\beta)\alpha^*\} = 2\omega_A^2|\lambda|^2|\beta|^2\,\text{Re}\{i/\Delta\omega_\pm\},\\ \frac{d}{dt'}|\beta|^2 &= 2\,\text{Re}\{(-i\lambda^*\omega_B\alpha)\beta^*\} = 2\omega_B^2|\lambda|^2|\alpha|^2\,\text{Re}\{i/\Delta\omega_\pm\}.\end{aligned} \tag{17}$$

In the second identity, we used $\alpha = \beta\lambda\omega_A/\Delta\omega_\pm$ and $\beta = \alpha\lambda^*\omega_B/\Delta\omega_\pm$, which holds true for the eigenmodes (here $\Delta\omega_\pm = \omega'_\pm - \omega'_c = \pm|\lambda|\sqrt{\omega_A\omega_B}$.). There is a striking resemblance between Eq. (17) and the model (2) in Sect. 2. Clearly, we can identify $G_{AB} = 2\omega_A^2|\lambda|^2\,\text{Re}\{i/\Delta\omega_\pm\}$ and $G_{BA} = 2\omega_B^2|\lambda|^2\,\text{Re}\{i/\Delta\omega_\pm\}$ with the symbol correspondence $1 \to A$ and $2 \to B$. Note that the sign of the coefficients depends on the mode $\pm$. When the system is stable, corresponding to an interaction between modes with the same frequency sign, both coefficients vanish:

$G_{AB} = G_{BA} = 0$. In contrast, when the system is unstable $G_{AB}G_{BA} = 4|\omega_A \omega_B||\lambda|^2 > 0$, in agreement with the discussion of Sect. 2.

This connection between the two models sheds light into the physics underlying the unstable interactions between modes with positive and negative frequencies. As already noted in Sect. 2, when $G_{AB}G_{BA} > 0$, or equivalently $\text{Im}\{\omega_\pm\} \neq 0$, either the modes provide simultaneously power to each other ($G_{AB}, G_{BA} > 0$), leading to an exponential growth, or they absorb energy from one another ($G_{AB}, G_{BA} < 0$), leading to an exponential decay. Hence, the imaginary part of the perturbed frequency is associated with a flow of energy from the source that drives the time modulation to the crystal or the other way around. Previous works have discussed similar instabilities as a result from coherently coupling a positive-mass harmonic oscillator with a negative-mass oscillator [39-41] and in moving-type platforms [59-67].

The particle-hole interactions are always present in our system due to the flatness of the low-frequency bands. In fact, in dispersive media, the phase velocity $v_{ph} = \omega/k$ (and also the group velocity) of some frequency bands may become arbitrarily small for high wavenumbers, i.e., $v_{ph} \to 0$ as $k \to \infty$ (here, we ignore possible nonlocal effects). Therefore, a traveling-wave modulation of a dispersive system is always operating in the superluminal regime, no matter how small the modulation velocity is [67]. In Sect. 3.4, we show that for small $v$ the instabilities are more sensitive to the effect of material loss. In a quantum mechanical description the particle-hole instabilities are associated with photon pair production.

It is useful to summarize the previous analysis regarding the effect of weak spacetime perturbations by highlighting the following results:

1) The traveling-wave modulation creates resonant couplings between unperturbed modes whose wavevectors $k_{A/B}$ and frequencies $\omega_{A/B}$ in the lab frame satisfy the "selection rules" (8):

$$k_A - k_B = N\frac{2\pi}{a}, \quad N \in \mathbb{Z}, \quad \text{and} \quad \omega_A - vk_A = \omega_B - vk_B. \qquad (18)$$

2) In the absence of intrinsic loss ($\Gamma_A = \Gamma_B = 0$), the interaction gives rise to particle-hole instabilities when

$$\omega_A \omega_B < 0. \qquad (19)$$

Otherwise, the mode hybridization originates a conventional band gap for $\lambda \neq 0$.

### *3.4 Loss as a stabilizing mechanism*

So far, we have used a model that describes an idealized linear system without intrinsic loss. Next, we consider the inevitable Joule-dissipation that arises in realistic materials ($\Gamma \neq 0$). For simplicity, it is assumed that the "collision frequency" $\Gamma$ is independent of space and time. The two-band model developed in Sect. 3.2 can be readily extended to the lossy case. In fact, the full Hamiltonian is now of the form $\hat{H}_{co} = \hat{H}_{co,0} + \hat{H}_{pert,tot}$, where $\hat{H}_{co,0}$ is still the operator that describes the continuum lossless material in the primed coordinates, and $\hat{H}_{pert,tot} = \hat{H}_{pert1} + \hat{H}_{pert2}$ includes the perturbations due to the modulation ($\hat{H}_{pert1}$), already studied in Sect. 3.2, and the perturbation due to loss $\hat{H}_{pert2}$:

$$\hat{H}_{\text{pert2}} = -i\Gamma \begin{pmatrix} 0 & 0 & 0 & 0 \\ 0 & 0 & 0 & 0 \\ 0 & 0 & 0 & 0 \\ 0 & 0 & 0 & 1 \end{pmatrix}. \tag{20}$$

Straightforward calculations show that $\langle \psi_n | \hat{H}_{\text{pert2}} | \psi_m \rangle = \left\{ \dfrac{-i\Gamma}{\varepsilon_0 \omega_p^2} e^{i(N_m - N_n)\frac{2\pi}{a}x'} \right\}_{\text{av}} \omega_n \omega_m \dfrac{P_n^* P_m}{W_0}$.

Since $\Gamma$ is a constant, only the $n = m$ terms are nonzero: $\langle \psi_n | \hat{H}_{\text{pert2}} | \psi_m \rangle = -i\gamma_n \delta_{n,m}$, where $\gamma_n = \dfrac{\Gamma}{\varepsilon_0 \omega_p^2} \omega_n^2 \dfrac{|P_n|^2}{W_0}$. Thus, the Hamiltonian is now represented by the matrix:

$$\hat{H}_{\text{co}} \to \begin{pmatrix} \omega_c' - i\gamma_A & \lambda \omega_A \\ \lambda^* \omega_B & \omega_c' - i\gamma_B \end{pmatrix}. \tag{21}$$

The perturbed frequencies in the co-moving frame must be updated to

$$\omega_\pm' = \omega_c' - i\frac{\gamma_A + \gamma_B}{2} \pm \frac{1}{2}\sqrt{4|\lambda|^2 \omega_A \omega_B - (\gamma_A - \gamma_B)^2}. \tag{22}$$

As a result, the instability condition ($\text{Im}\{\omega_-'\} > 0$ or $\text{Im}\{\omega_+'\} > 0$) becomes:

$$-\omega_A \omega_B > \omega_T^2, \quad \text{with} \quad \omega_T = \frac{\sqrt{\gamma_A \gamma_B}}{|\lambda|} = \frac{\Gamma \omega_A \omega_B}{\left| \left\{ \delta\tilde{\omega}_0^2 e^{i(N_B - N_A)\frac{2\pi}{a}x'} \right\}_{\text{av}} \right|}. \tag{23}$$

The instability threshold is controlled by the parameter $\omega_T$. Clearly, the intrinsic losses in the material act as a stabilizing mechanism that counterbalances the unstable particle-hole resonances as it increases the threshold $\omega_T$. From a different perspective, when $\Gamma$ is nonzero, the real part of the dispersion diagram is not any longer flat, and thus the phase velocity is nontrivial. Thus, the minimum modulation speed to operate in the super-luminal regime becomes nonzero.

### 3.5 Numerical Study

To validate the previous analysis based on perturbation theory, we compare the results predicted from Eq. (22) with the band structure determined using an exact formulation.

Figure 4 represents the exact band structure in the Galilean co-moving frame of the dispersive spacetime crystal near the band crossings marked in Fig. 2d for a dissipative perturbation such that $\Gamma = 0.01\omega_p$. The numerical results were computed using a transmission matrix approach [42].

The comparison between the exact and the approximate results shows that the developed perturbation analysis correctly predicts the particle-hole instability at the black crossing (a) between modes with positive and negative frequencies, as well as the stable band gap at the white crossing (b) between modes with positive frequencies. The quantitative correlation between the exact results and the perturbation theory is remarkably good. It is important to note that the spectrum is slightly displaced towards the lower-half frequency plane, a shift attributable to the loss effect disrupting the parity-time symmetry of the system.

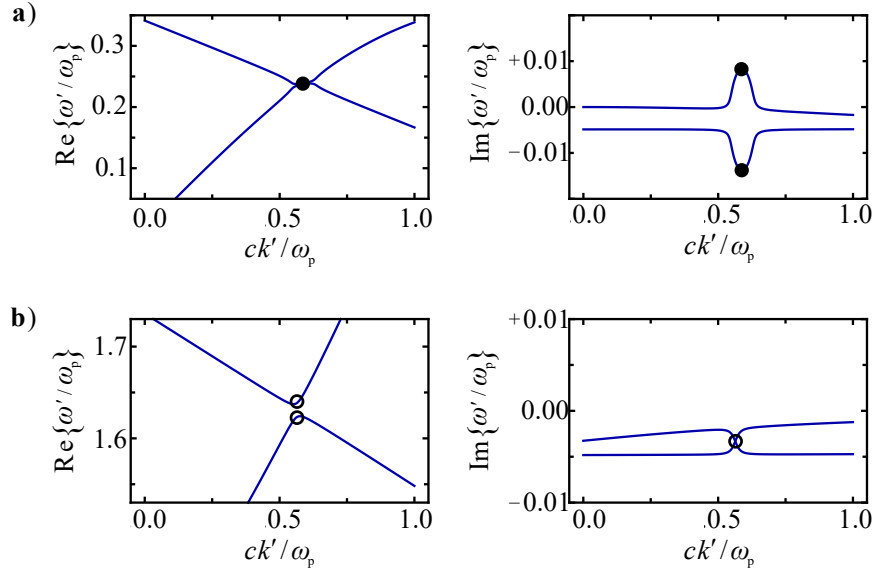

**Fig. 4** Exact (blue) band structure of the dispersive spacetime crystal with a binary structure near the black (a) and white (b) band crossings shown in Fig. 2d for the modulation speed $v = 0.18c$ and collision frequency $\Gamma = 0.01\omega_p$. The remaining structural parameters are as in Fig. 3. We represent the approximate perturbative results using the same black and white circles as in Fig. 2d.

## 4. Conclusions

We discussed how periodic time-varying modulations couple modes in different frequency bands: while the interaction between modes with the same frequency sign is usually stable, the interaction between modes with frequencies of opposite sign can lead to negative-mass-like instabilities in the system. In particular, we distinguish the second-order unstable interactions that may result from this coupling, from the case of parametric resonances that describe first-order feedback loops between modes linked by the particle-hole symmetry of the electromagnetic spectrum.

We studied the stability of a spacetime crystal described by a time-varying Drude-Lorentz model with a traveling-wave modulation. Using perturbation theory, we have shown that the coupling between modes of positive and negative frequency bands induces instabilities governed by specific "selection rules". These instabilities are most pronounced in modes exhibiting particle-hole duality. Moreover, our study underscores that any spacetime modulation in Lorentz-dispersive materials, for arbitrarily small modulation speeds, invariably correlates with a superluminal regime, a domain where instabilities are inevitably observed. To validate the findings of our two-band perturbation analysis, we conducted a successful comparison with the system's exact band structure. Our study offers practical insights into how efficiently extracting gain from time-variant systems, with potential applications in photonic circuits.

**Funding.** Institution of Engineering and Technology (IET); Simons Foundation (Award No. 733700); Instituto de Telecomunicações (Project No. UIDB/50008/2020) and Fundação para a Ciência e a Tecnologia under projects 2022.06797.PTDC. E. G. acknowledges funding from the Simons Foundation through a Junior Fellowship of the Simons Society of Fellows (855344, EG). P. A. H. acknowledges financial support from the Spanish Ministry of

Science and Innovation through the Ramón y Cajal program (Grant No. RYC2021-031568-I) and through Project No. PID2022-141036NA-I00 financed by MCIN/AEI/10.13039/501100011033 and FSE+.